\documentclass[journal]{IEEEtran}
\usepackage{amsmath,amsfonts}
\usepackage{algorithmic}
\usepackage{algorithm}
\usepackage{array}
\usepackage{textcomp}
\usepackage{stfloats}
\usepackage{url}
\usepackage{verbatim}
\usepackage{graphicx}
\usepackage{subfigure}
\usepackage{cite}
\usepackage{multicol}
\usepackage{color}
\usepackage{multirow}

\newcommand{\blue}[1]{}

\begin{document}
\title{An Entropy-Awareness Meta-Learning Method for SAR Open-Set ATR}
%
%
%

\author{Chenwei Wang,~\IEEEmembership{Student Member,~IEEE,}
        Siyi Luo, 
        Jifang Pei,~\IEEEmembership{Member,~IEEE,}
        Xiaoyu Liu,~\IEEEmembership{Student Member,~IEEE,}
        Yulin Huang,~\IEEEmembership{Senior Member,~IEEE,}
        Yin Zhang,~\IEEEmembership{Member,~IEEE,}
        and Jianyu Yang,~\IEEEmembership{Member,~IEEE}
\thanks{This work was supported by the National Natural Science Foundation of China under Grants 61901091 and 61901090.(\emph{Corresponding author: Jifang Pei.)}}
\thanks{The authors are with the Department of Electrical Engineering, University of Electronic Science and Technology of China, Chengdu 611731, China (e-mail: peijfstudy@126.com; dbw181101@163.com).}}

\maketitle

\begin{abstract}
Existing synthetic aperture radar automatic target recognition (SAR ATR) methods have been effective for the classification of seen target classes. However, it is more meaningful and challenging to distinguish the unseen target classes, i.e., open set recognition (OSR) problem, which is an urgent problem for the practical SAR ATR. The key solution of OSR is to effectively establish the exclusiveness of feature distribution of known classes. In this letter, we propose an entropy-awareness meta-learning method that improves the exclusiveness of feature distribution of known classes which means our method is effective for not only classifying the seen classes but also encountering the unseen other classes. Through meta-learning tasks, the proposed method learns to construct a feature space of the dynamic-assigned known classes. This feature space is required by the tasks to reject all other classes not belonging to the known classes. At the same time, the proposed entropy-awareness loss helps the model to enhance the feature space with effective and robust discrimination between the known and unknown classes. Therefore, our method can construct a dynamic feature space with discrimination between the known and unknown classes to simultaneously classify the dynamic-assigned known classes and reject the unknown classes. Experiments conducted on the moving and stationary target acquisition and recognition (MSTAR) dataset have shown the effectiveness of our method for SAR OSR.
\end{abstract}


%
\IEEEpeerreviewmaketitle

\section{Introduction}
\IEEEPARstart{A}s an essential microwave remote sensing system, synthetic aperture radar (SAR) is increasingly applied in both the civil and military fields, while SAR automatic target recognition (ATR) is one of the most important points in SAR applications. With the continuous development in recent years, many SAR ATR methods have been proposed and good results have been achieved as well \cite{intro2,wang2023entropy,wang2022recognition,wang2023sar,wang2022global,wang2022semi,wang2020deep,intro3,intro5}.

In applications, it is more interesting and challenging to distinguish the unseen target classes, i.e., {the} open set recognition (OSR) problem. 
Besides, when there are some targets with unknown classes, most existing SAR ATR methods will assign these targets as known classes, which leads to unacceptable {mistakes} and potential risks. 
For example, it may lead to the collapse of battlefield reconnaissance, when some unseen dangerous classes are classified as known safety classes.
Furthermore, no matter how large the training dataset is, there are always unknown target classes in real-world SAR ATR missions. Therefore, a SAR ATR method should have the capabilities for OSR, i.e., identifying the unknown classes besides recognizing the known classes.

The goal of the OSR is to build a model which can not only recognize the known classes but also reject unknown class targets as outliers without any information about the unknown classes \cite{introexample1}. It is much closer to the practical scenes and has a similar self-awareness to that of humans to identify what is familiar and refuse what is unfamiliar. 

The OSR problem is crucial and fundamental in real-world SAR ATR and has been noticed for a long time \cite{introexample1,wang2020deep,wang2022sar,wang2021multiview,wang2019parking,wang2021deep, introexample2}. The basis and importance of the OSR have promoted the research in SAR ATR \cite{introexample3}. For example, Dang \textit{et al.} \cite{introexample4} used the class boundaries to train the open set recognition/outlier detection model and their proposed O-SAR method works well on the accuracy of false target rejection. 

However, even though these existing methods have shed light on the OSR in SAR ATR, they are mainly based on hand-crafted features or prior knowledge, and the recognition performance mainly depends on the quality of this prior information. Furthermore, they also need to set a hand-crafted threshold value to discriminate between the known and unknown classes. Most of them just construct the static feature space based on the whole closed dataset and design hand-craft features to reject limited open classes, which is not always effective for the problem of the OSR in practice.

Essentially, the key solution of OSR is to effectively establish the exclusiveness of feature distribution of known classes. With the exclusiveness of feature distribution, the  feature space can discriminate between the known and unknown classes and enhance the effectiveness of the discrimination.
In light of the vigorous development and superior performance of meta-learning, we focus on the issue of “learning to learn” by a meta-learner \cite{metasurvey}. 
Different from the traditional mini-batch training which construct the feature space based on the training dataset, meta-learning can constructs the feature space based on the formulation of the tasks which are the basic units of the meta-learning. There are some meta-based methods proposed to validate the effectiveness of SAR ATR or few-shot SAR ATR \cite{meta-learning2,wang2020multi,li2023panoptic,liang2023efficient,meta-learning3}.

Therefore, based on the principle of meta-learning, we propose an entropy-awareness meta-learning method for SAR open set ATR. By dynamically assigning the known and unknown classes, our method trains the model to construct the feature space based on the dynamic-assigned known classes and forces the feature space to discriminate between the unknown and known classes. Furthermore, the entropy-awareness loss for the OSR problem is proposed to help the model to enhance the feature space with effective and robust discrimination between the known and unknown classes.
The main contributions of this letter are as follows.

1) We propose a meta-learning framework for the OSR problem in SAR ATR. By ceaselessly facing the dynamic-assigned known classes and rejecting the dynamic-assigned unknown classes, the framework can construct a robust feature space that provides discrimination between the known and unknown classes to solve the problem of OSR.

2) The entropy-awareness loss is proposed for the recognition and rejection, consisting of the meta cross-entropy loss, the entropy-distancing loss, and the open-set loss. The three parts of entropy-awareness loss aim to maximize the entropy of the feature distances of the known classes, enlarge the feature distances between the known and unknown classes, and enhance the effectiveness of the discrimination between the known and unknown classes, respectively.

3) The proposed method achieves high performance on the standard benchmark. The recognition accuracy of the known classes is superior and the recall and precision have achieved excellent levels.

The remaining structure of this paper is organized as follows. The proposed method is described in Section \uppercase\expandafter{\romannumeral2}. Section \uppercase\expandafter{\romannumeral3} illustrates the experiments and results. Finally, the conclusion is drawn in Section \uppercase\expandafter{\romannumeral4}.

\begin{figure*}
\centering
\includegraphics[width=0.78\textwidth]{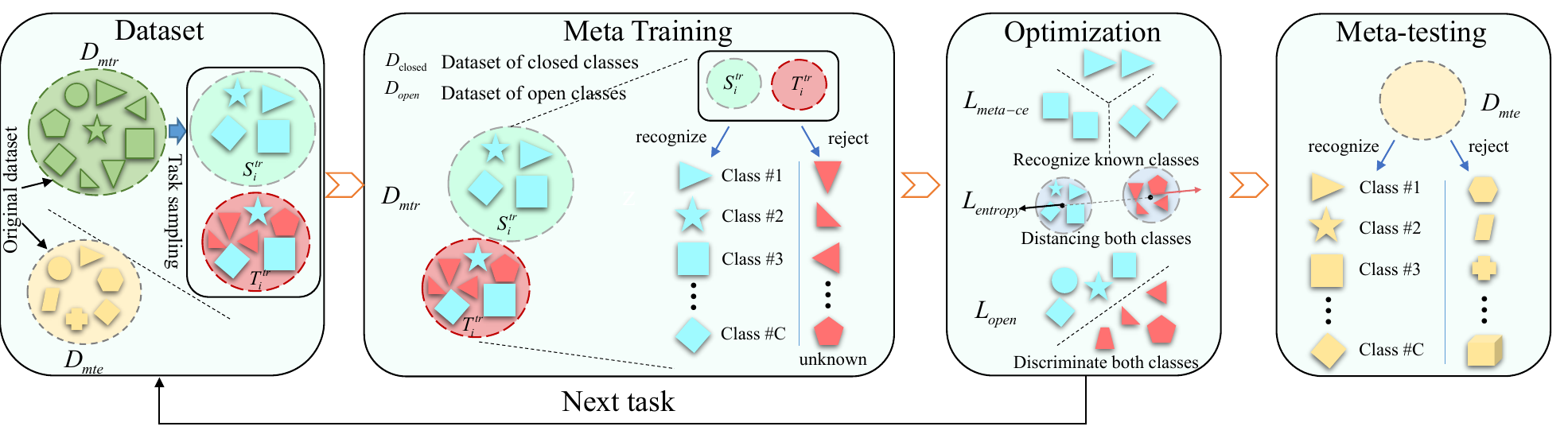}
\caption{Whole framework of proposed method. Dataset is first divided into a meta training set and a meta testing set. In the meta training phase, the meta training set is used for task-based meta-training. Entropy-awareness loss is for optimization consisting of three parts.}
\label{framework}
\end{figure*}

\section{Proposed Method}
In this section, the problem formulation of the OSR in SAR ATR is described. Then the proposed method is introduced in detail, including the meta-learning framework for OSR and entropy awareness loss. 

\subsection{Problem Formulation}

\begin{figure*}
\centering
\includegraphics[width=0.76\textwidth]{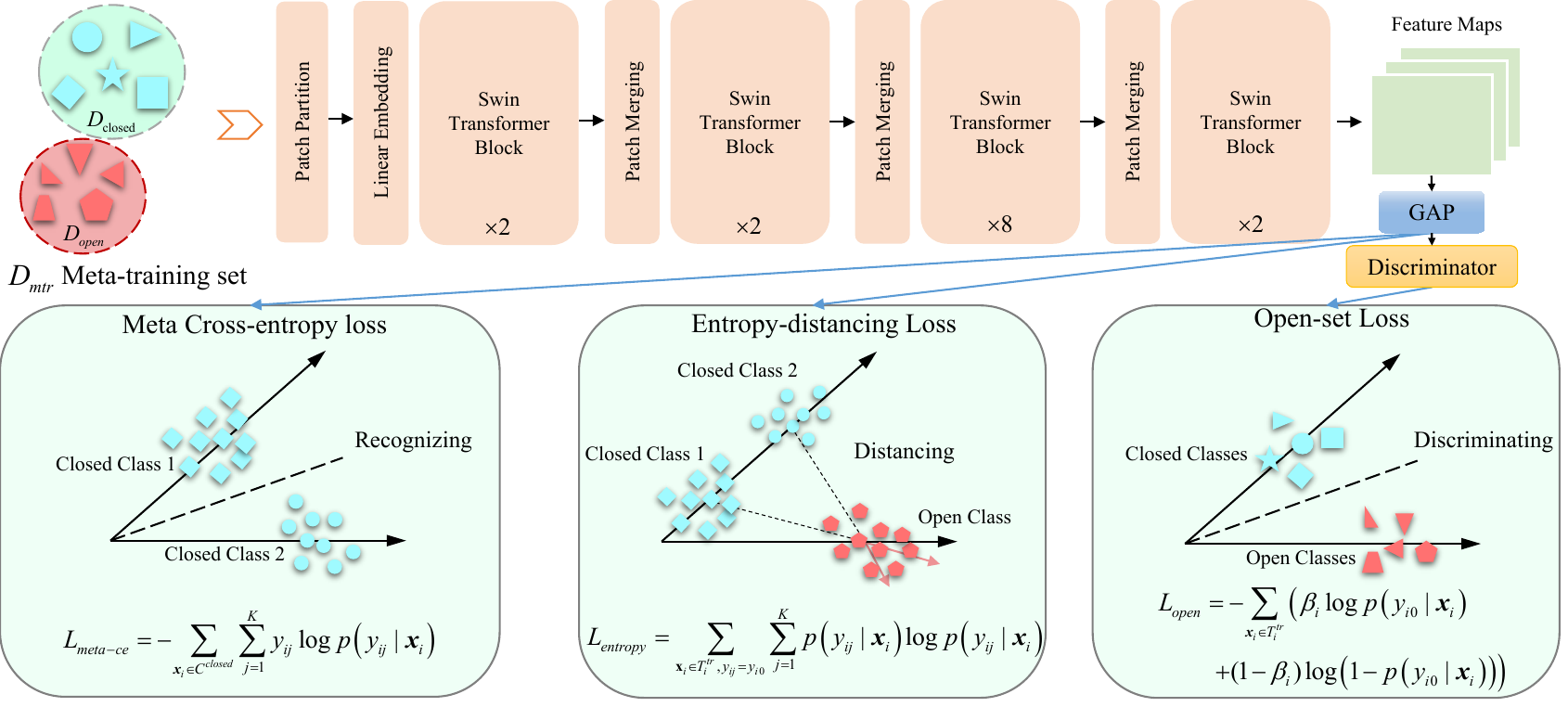}
\caption{Structure of proposed method and details of three parts of the entropy-awareness loss. Meta cross-entropy loss is for the recognition of the known. Entropy-distancing loss maximizes the posterior entropy and detects the unknown classes. Open-set loss is for discriminating between the known and unknown classes. GAP means the global average pooling layer. The yellow discriminator is one dense layer with SoftMax.}
\label{method}
\end{figure*}

The OSR problem refers to an open-set scenario that has some novel class types in testing which never existed in training. Given a labeled training set for the model ${D_{train}} = \left\{ {\left( {{\mathbf{x}_i},{y_{ij}}} \right)} \right\}_{i = 1}^N \subset C$. $C$ is the set of the known classes, and ${\mathbf{x}_i}$ is the $i$th SAR image, $y_{ij}$ is the label of ${\mathbf{x}_i}$, mean ${\mathbf{x}_i}$ belongs to ${j}$th class. In the closed set scenario, the model is evaluated on a testing set in which the labels are also drawn from the same set of classes ${D_{test - closed}} = \left\{ {\left( {{\mathbf{x}_i},{y_{ij}}} \right)} \right\}_{i = 1}^M \subset C$. Meanwhile, the model will give the prediction over the known class as $ p\left( {y\left| {\bf{x}} \right.} \right)$. The classes of the testing SAR images will be assigned to the class with the maximum probability in the probability vector.

For the OSR, given an unknown testing set ${D_{unknown}} = \left\{ {\left( {{\mathbf{x}_i},{y_{i0}}} \right)} \right\}_{i = 1}^V \subset U$. Here, $U$ is the set of unknown classes. In the testing phase, the testing images may also come from the unknown testing set ${D_{unknown}}$. Therefore, the testing dataset can be presented as $ {D_{test - open}} = \left\{ {\left( {{\mathbf{x}_i},{y_{ij}}} \right)} \right\}_{i = 1}^{V+M} \subset \left( {C \cup U} \right)$. In addition, the model not only predicts a probability distribution over the known class but also gives a score or value to predict whether the testing image belongs to any of the known classes.

After the definition of the OSR problem, the process of meta-learning will be employed to make the model classify the known classes and reject the unknown classes precisely.

\subsection{Meta-learning Framework for OSR}
The pipeline of the proposed method is shown in Fig.\ref{framework} and consists of two main phases: meta-training and meta-testing. In the meta-training, by randomly selecting the known and unknown classes per episode, the meta-learner faces many different tasks and is optimized by the performance on these tasks, finally having the capability of recognizing the known classes and rejecting unknown classes. In the phases of the meta-testing, the performance of the OSR problem is evaluated.

Regarding the tasks as the fundamental unit in the training and testing phases,
there are also one meta training dataset $D_{mtr}$ and one meta testing dataset $D_{mte}$ in the process, as shown in Fig.\ref{framework}. 
The meta-training dataset $D_{mtr}$ is regarded as a closed dataset in practical application. 
Then the model is trained on the closed dataset $D_{mtr}$ and expected to be able to recognize the class types in the closed dataset and reject the class types out of the closed dataset in the testing phase.
The details of the meta-training and meta-testing phases are as follows.

For all tasks sampled in the meta-training phase, the dataset can be formulated as $D_{mtr}=\left \{ \left ( S_i^{tr},T_i^{tr} \right )  \right \} ^{N^{tr}}\subset C$, where $C$ means the collection of known classes. There is one training subset $S_i^{tr}$ and one testing subset $T_i^{tr}$ for the $i{\rm{th}}$ task, where $N^{tr}$ is the number of all the tasks in the meta training phase. For every task, given the 'known classes' $C^{closed}$ and the 'unknown classes' $U^{open}$, the two subsets meet two requirements: $C^{closed}\cup U^{open}=C$ and $C^{closed}\cap U^{open}=\emptyset $. The training subset $S_i^{tr}$ is sampled from the known classes $C^{closed}$, and the testing subset $T_i^{tr}$ is sampled from all the classes $C=C^{closed}\cup U^{open}$.

For the $i{\rm{th}}$ task, the meta-training phase can be divided into two steps. 
Given the meta learner $f_\theta $ and the suitable loss $L$, $\theta $ is the parameter of the meta learner, the first step is to optimize on $S_i^{tr}$ as
\begin{equation}
f_\theta ^{'}{\text{ = }}\arg \min L\left( {y_S^{tr},{S^{tr}}} \right)
\end{equation}
where $y_S^{tr}$ is the corresponding label of $S^{tr}$, ${f_\theta ^{'}}$ is the optimal meta learner under the whole training subset $S^{tr}$. 

The second step is to optimize on $T^{tr}$ and  find the meta learner
\begin{equation}
{f_\theta ^{*}}{\text{ = }}\arg \min L\left ( y_T^{tr},T^{tr} \right ) 
\end{equation}
where $y_T^{tr}$ is the corresponding label of $T^{tr}$, ${f_\theta ^{*}}$ is returned as the optimal meta learner for the $i{\rm{th}}$ task. Through the suitable loss and optimization procedure, the optimal meta learner ${f_\theta ^{*}}$ for the $i{\rm{th}}$ task can be optimized and input as the initial meta learner for the $(i+1){\rm{th}}$ task.

After the meta learner is optimized under $N^{tr}$ meta training tasks, the meta learner will be evaluated in the meta-testing phase. 
The meta testing dataset is also formulated as $D_{mte}=\left \{ \left ( S_i^{te},T_i^{te} \right )  \right \} ^{N^{te}}\subset \left (C\cup U\right) $. The training subset $S^{te}$ is sampled from the class types of the closed dataset $C$, and the testing subset $T^{te}$ is sampled from all the class types $\left( C\cup U \right)$.
The training subset $S^{te}$ is used by the meta learner ${f_\theta ^{*}}$ to generate the optimal model
\begin{equation}
{f_\theta ^{''}}{\text{ = }}\arg \min L\left ( y_S^{te},S^{te} \right ) 
\end{equation}
where $y_S^{te}$ is the corresponding label of $S^{te}$. The performance of the method will be evaluated using the final meta learner ${f_\theta ^{''}}$ on $T^{te}$.

In this way, the model can construct the feature space based on the dynamic closed classes, and the feature space has the capability of rejecting all other classes not belonging to the dynamic closed classes.
Then, the entropy-awareness loss is described in details.

\subsection{Entropy-Awareness Loss for OSR}

As shown in Fig. \ref{method}, the entropy-awareness loss consists of three parts, the meta cross-entropy loss, the entropy-distancing loss, and the open-set loss. 
The meta cross-entropy loss helps the model to construct a basic feature space by maximizing the entropy of the feature distances of the known classes and provides the basic ability to recognize the known classes. 
The entropy-distancing loss calculates and reduces the entropy of the unknown classes assigned to the known classes to enlarge the feature distances between the known and unknown classes. 
The open-set loss aims to force the constructed feature space to obtain discrimination between the known and unknown classes. 
In one word, three losses help the model to provide effective and robust discrimination between the known and unknown classes and discriminate them in the feature space with precise recognition of the known classes.

Given the meta training dataset $D_{mtr}=\left \{ \left ( S_i^{tr},T_i^{tr} \right )  \right \} ^{N^{tr}}\subset C$ and the meta learner ${f_\theta ^{'}}$ trained on the $S^{tr}$, the entropy-awareness loss can be presented as 
\begin{equation}
L_{ea}=\lambda _{1}L_{meta-ce}+\lambda _{2}L_{entropy}+\lambda _{3}L_{open}
\end{equation}
where $L_{meta-ce}$ means the meta cross-entropy loss, $L_{entropy}$ is the entropy-distancing loss,
$L_{open}$ is the open-set loss, $\lambda _{1}$, $\lambda _{2}$ and $\lambda _{3}$ are the weighing parameters. 
The meta cross-entropy loss $L_{meta-ce}$ is based on the $T^{tr}$ and calculated as
\begin{equation}
L_{meta-ce}=-\sum_{{\bf{x}}_{i} \in C^{closed}} \sum_{j=1}^K y_{ij}\log{p\left ( y_{ij} | \mathbf{x}_i  \right ) } 
\end{equation} 
where $C^{closed}$ is the known classes for the task $T_i^{tr}$, $K$ is the known class number in $C^{closed}$, $\left ( {\bf{x}}_i,y_{ij} \right ) $ belongs to the close-set in $T_i^{tr}$. $p\left( {{y}_{ij}}|{{\bf{x}}_{i}} \right)=sf\left( {\text{L}_{2}}\left( -d\left( \mathbf{v}\left( {{x}_{i}} \right),{{\mathbf{a}}_{{{j}}}} \right) \right) \right)$, ${\text{L}_2}\left(  \cdot  \right)$ is the $\text{L}_2$ normalization, $sf$ is the SoftMax function,
${\mathbf{v}}\left( {{\mathbf{x}_i}} \right)$ is the final feature maps of $\mathbf{x}_i $, ${{\mathbf{a}}_{{{j}}}}$ is the prototype of the $j$th class, the calculation of ${{\mathbf{a}}_{{j}}}$ is the same as in \cite{prototypical}. The meta cross-entropy is designed to make the meta learner recognize the known classes.

Then, for ${\bf{x}}_i$ belongs to the open-set in $T_i^{tr}$, the entropy-distancing loss $L_{entropy}$ can be presented as
\begin{equation}
L_{entropy}=\sum_{{\mathbf{x}_i}\in T_i^{tr}, y_{ij}=y_{i0}} \sum_{j=1}^K p\left ( y_{ij} | {\mathbf{x}_i} \right ) \log{p\left ( y_{ij} | {\mathbf{x}_i}  \right ) } 
\end{equation}
where $y_{i0}$ indicates the corresponding sample ${\mathbf{x}_i}$ belongs to the open set, while 
$j\ge 1$ indicates that the sample is in the closed set.

The entropy-distancing loss is designed to minimize the entropy of the feature distances of the unknown classes over the known classes and make the meta-learner reject the unknown classes. The features go through the discriminator which is one dense layer with SoftMax, thus the open-set loss is calculated as
\begin{equation}
\begin{aligned}
  {{L}_{open}}=-\sum\limits_{ {{\mathbf{x}}_{i}}\in T_{i}^{tr}} & {\left( {{\beta }_{i}}\log p\left( {{y}_{i0}}|{{\mathbf{x}}_{i}} \right) \right.} \\ 
 & \left. +(1-{{\beta }_{i}})\log \left( 1-p\left( {{y}_{i0}}|{{\mathbf{x}}_{i}} \right) \right) \right)  
\end{aligned}
\end{equation}
where ${p\left ( y_{i0} | {\mathbf{x}_i}  \right ) }$ means the probability that ${\mathbf{x}_i}$ is predicted to be in the open set, and $\beta_i=1$ if $x_{i}$ belongs to the open set, else $\beta_i=0$. When facing unknown classes, the meta learner should not assign it to any known class with a large probability. Through the three parts of the entropy-awareness loss, the meta-learner can help the discriminator to reject the unknown class when the entropy of feature distance is large. Finally, the precise recognition among the known classes and rejection for the unknown classes can be achieved. 

\begin{figure}[htb]
\centering
\centering\includegraphics[width=0.38\textwidth]{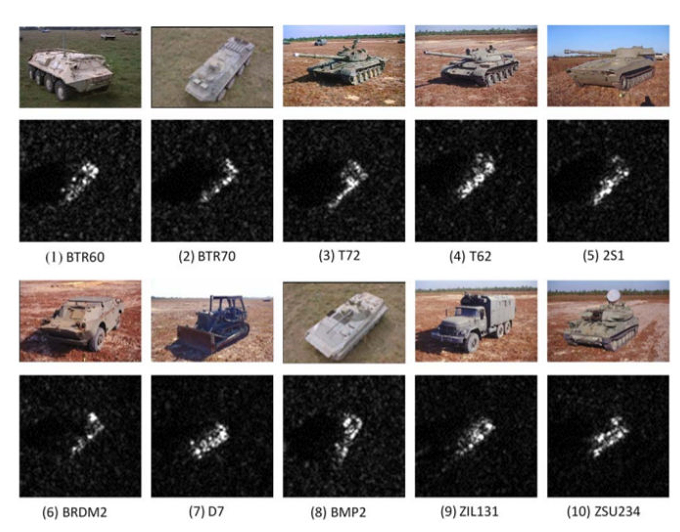}
\caption{Optical images and corresponding SAR images of ten classes of objects in MSTAR dataset.}
\label{dataset}
\end{figure}

\section{Experimental Results}
In this section, the benchmark dataset is introduced, and the employed metrics for the OSR performance are presented. The experiments and results are presented in detail. 
\subsection{Dataset and Meta Training Settings}
The MSTAR dataset is a benchmark dataset for the SAR ATR performance assessment. The dataset contains a series of $0.3m\times0.3m$ SAR images of ten different classes of ground targets. The optical images and corresponding SAR images of ten classes of targets in the MSTAR dataset are shown in Fig. \ref{dataset}. 
We choose six classes as the known classes and the rest of the classes as the unknown classes. For each episode in the meta-training phase, we randomly choose 4 classes from the 6 known classes and set the rest of the 2 classes as the unknown classes. The size of input SAR images is $224 \times 224$ by applying bilinear interpolation to the raw data. The sample numbers for the support, query, and open samples in one episode are all set as 10. The values of $\lambda _{1}$, $\lambda _{2}$ and $\lambda _{3}$ are set as 0.5, 0.25 and 0.25. The learning rate is initialized as 0.01 and reduced with the 0.5 ratios for every 1000 episodes. 

\subsection{Metrics and Performance of Recognition and Rejection}
For quantitive evaluation of the performance of the OSR in SAR ATR, four metrics are employed, i.e. true-positive rate (TPR), false-positive rate (FPR), recall, and precision. 

In the experiment of standard operating condition (SOC), the training images of the known classes under 17\textdegree are set as the meta-training dataset and the support set of the meta-testing set, and the testing images under 15\textdegree are set as the meta-testing dataset. In the experiments of extended operating condition depression-variant (EOC-D) and version-variant (EOC-VV), the meta-training dataset is the same as in SOC.

The recognition performances and rejections for the known and unknown classes under SOC and EOCs are presented in Table \ref{recognitionperformance}. It is clear that our method obtains precise rejection for the unknown classes from the results of FPR, and recognizes the known classes precisely from the results of TPR and recall. 

From the experiments of the OSR, it is clear that our method obtains the superior recognition performance and the OSR performance under different training samples.
The recognition ratios of EOC-D and EOC-VV, the OSR performances under different training samples achieves high performance. 

From the performances of SOC and EOCs, the results have illustrated that our method has robustness and effectiveness for the OSR.

\subsection{Comparison, Ablation and Feature Visualization}

In this section, the comparison with other OSR methods and the ablation experiments are presented. The feature visualizations corresponding the ablation experiments are also shown.

The comparison methods are OpenMAX \cite{openmax}, EVM \cite{EVM}, OsmIL \cite{OSmIL}, O-SAR \cite{o-sar} , and ML-GAN \cite{ml-gan}. The thresholds for OpenMAX, EVM, OsmIL, O-SAR, and ML-GAN are set to 0.20, 0.50, 0.01, 0.0001, and 0.28, respectively. The experiments of the comparison method and our method are run under the conditions that 3 classes, BMP2, BTR70 and T72, are the known classes, the resting 7 classes are the unknown classes. 

\begin{figure}[htb]
\centering
\centering\includegraphics[width=0.48\textwidth]{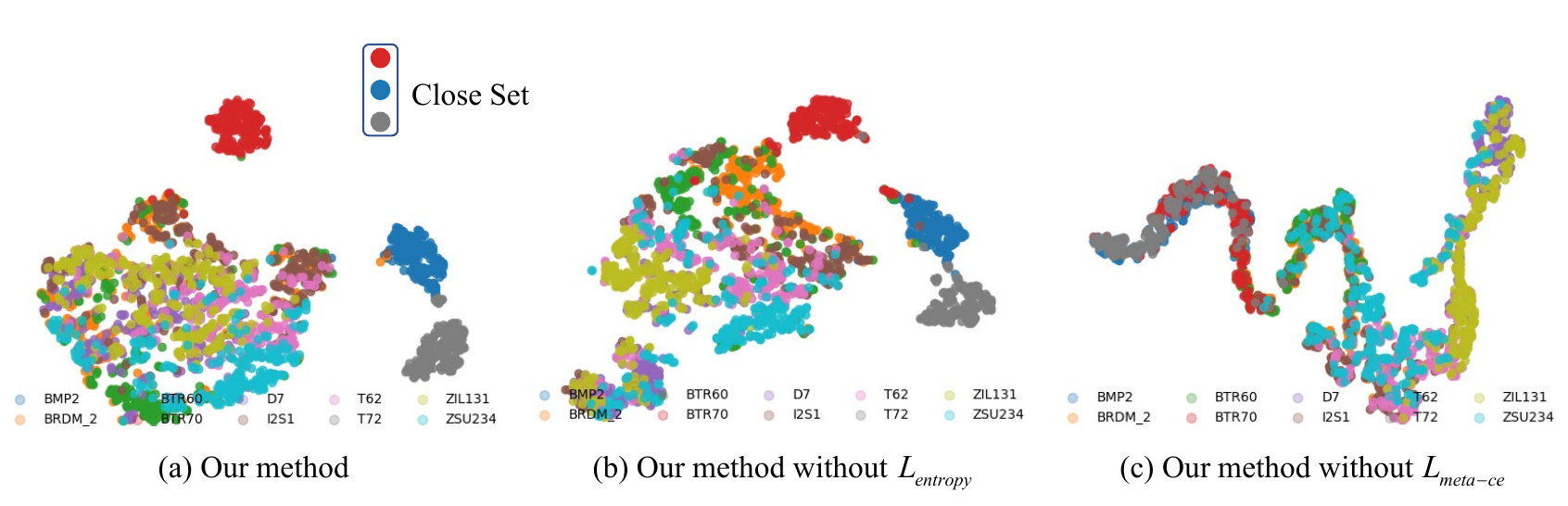}
\caption{Feature distributions of our methods under different configurations. (a) full version of our methods. (b) our method without $L_{entropy}$, (c) our method without $L_{meta-ce}$.}
\label{ablation}
\end{figure}

The comparison with other state-of-art OSR methods is shown in Table \ref{comparison}. It is clear that our method can achieve better OSR performance from the comparison of four metrics, TPR, FPR, recall and precision. At the same time, our method also obtain the best recognition performance. The ablation experiments are also shown in Table \ref{comparison}. We run three ablation experiments to validate the soundness of our method under the same experiment configuration as the comparison. The fist one is our method without the meta cross-entropy loss, Eq. 5. The second one is our method without the entropy-distance loss, Eq. 6. The last one is the full version of our method. The corresponding feature visualizations are shown in Fig. \ref{ablation}. From Table \ref{comparison} and Fig. \ref{ablation}, it is clear that our main innovations can improve the feature distribution and the OSR performance.

From the comparison and ablation experiments, they have illustrated that our method can achieve superior OSR performance.

\begin{table}[htb]
\renewcommand{\arraystretch}{1.2}
\setlength\tabcolsep{3pt}
\caption{Results of recognition and rejection under SOC and EOCs}
\centering
\label{recognitionperformance}
\begin{tabular}{lccccc}
\hline\hline
\multicolumn{6}{c}{SOC}                                                                                                                 \\ \hline
Metric           & 6-way 20        & 6-way 40        & 6-way 60        & 6-way 80       & 6-way 100       \\ \hline
TPR           & 89.84\% & 91.97\% & 92.68\% & 96.95\% & 97.87\%     \\ 
FPR            & 7.57\%  & 3.24\%  & 1.57\%  & 1.08\%  & 0.39\%       \\ 
recall      & 91.03\% & 92.81\% & 93.39\% & 97.26\% & 97.88\%     \\ 
precision     &  94.26\% & 97.52\% & 98.79\% & 99.20\% & 99.71\%     \\ 
Accuracy       &  75.42\% & 92.28\% & 93.64\% & 95.16\% & 98.48\%    \\ \hline
\multicolumn{6}{c}{EOC-D}                                                                                                               \\ \hline
Metric           & 6-way 20        & 6-way 40        & 6-way 60        & 6-way 80       & 6-way 100       \\ \hline
TPR           & 94.78\% & 95.05\% & 95.83\% & 96.00\% & 97.04\%     \\ 
FPR            & 11.46\% & 10.76\% & 10.42\% & 8.68\%  & 7.99\%        \\ 
recall        & 94.79\% & 94.96\% & 95.83\% & 96.00\% & 97.05\%      \\ 
precision      &  89.20\% & 89.31\% & 90.18\% & 91.69\% & 92.38\%     \\ 
Accuracy       & 87.34\% & 90.15\% & 94.19\% & 92.93\% & 89.07\%    \\ \hline
\multicolumn{6}{c}{EOC-VV}                                                                                                              \\ \hline
Metric           & 6-way 20        & 6-way 40        & 6-way 60        & 6-way 80       & 6-way 100       \\ \hline
TPR         & 89.85\% & 90.43\% & 90.90\% & 93.47\% & 94.63\%     \\ 
FPR                 & 4.46\%  & 1.22\%  & 1.11\%  & 0.96\%  & 0.88\%  \\ 
recall        & 89.85\% & 90.43\% & 90.90\% & 93.47\% & 94.63\%      \\ 
precision       & 86.42\% & 95.92\% & 96.29\% & 96.86\% & 97.13\%      \\
Accuracy        & 89.87\% & 89.81\% & 95.12\% & 94.26\% & 98.27\% \\ \hline\hline
\end{tabular}
\end{table}

\begin{table}[htb]
\renewcommand{\arraystretch}{1.2}
\setlength\tabcolsep{7.8pt}
\caption{Comparison Results and Ablation Experiments of Recognition and Rejection under 3-way SOC.(w/o means without)}
\centering
\label{comparison}
\begin{tabular}{cccccc}
\hline \hline 
Methods & TPR & FPR & recall & precision & Accuracy \\ \hline 
OpenMAX \cite{openmax} & 74.9 & 20.7 & 76.0 & 67.8 & 78.2 \\
EVM \cite{EVM} & 91.8 & 7.7 & 90.5 & 81.0 & 91.8 \\
OSmIL \cite{OSmIL} & 93.4 & 6.9 & 93.4 & 87.0 & 93.7 \\
O-SAR \cite{o-sar} & 94.4 & 3.4 & 94.4 & 89.9 & 96.0 \\
ML-GAN \cite{ml-gan} & 99.3 & 0.9 & 98.2 & 97.0 & 98.8 \\
Ours w/o Eq. 5 & 96.1 & 4.6 & 96.1 & 87.0 & - \\ 
Ours w/o Eq. 6 & 99.7 & 3.6 & 99.7 & 90.0 & 96.7 \\ 
Ours & 99.7 & 0.4 & 99.7 & 98.6 & 99.4 \\ \hline \hline 
\end{tabular}
\end{table}

\section{Conclusion}
The OSR problem is crucial and fundamental in SAR ATR, which is a more challenging open environment setting without any information about the unknown classes than the normal close-set recognition. This limitation inevitably leads to hard discrimination between the known and unknown classes in the feature space. The proposed method learned to construct a dynamic feature space based on the known classes through the constructed tasks. To provide discrimination between the unknown and known classes, the entropy-awareness loss forces the model to enhance the feature space with effective and robust discrimination between the known and unknown classes. Experimental results on the MSTAR dataset have validated the effectiveness and robustness of our method in OSR in SAR ATR. Under the SOC and EOCs, our method can handle the large changes in the depression angle and version series in SAR images.

\bibliographystyle{IEEEtran}
\bibliography{references}


%






\end{document}